# Charge distribution and magnetism in bilayer $La_3Ni_2O_7$: a hybrid functional study

Kateryna Foyevtsova
*Department of Physics and Astronomy, University of Notre Dame, Notre Dame, IN, USA*
*Stavropoulos Center for Complex Quantum Matter,*
*University of Notre Dame, Notre Dame, IN, USA*
*Department of Physics & Astronomy, University of British Columbia,*
*Vancouver, British Columbia V6T 1Z1, Canada and*
*Stewart Blusson Quantum Matter Institute, University of British Columbia, Vancouver, British Columbia V6T 1Z4, Canada*

Ilya Elfimov and George A. Sawatzky
*Department of Physics & Astronomy, University of British Columbia,*
*Vancouver, British Columbia V6T 1Z1, Canada and*
*Stewart Blusson Quantum Matter Institute, University of British Columbia, Vancouver, British Columbia V6T 1Z4, Canada*

An accurate understanding of the ground state electronic properties of $La_3Ni_2O_7$, a high-temperature superconductor under pressure, is key for unveiling the origin of its superconductivity. In this paper, we conduct a theoretical study of the electronic structure of the bilayer polymorph of $La_3Ni_2O_7$ using the hybrid functional approach, which is well suited to tackle the non-local correlation effects arising in this system from the molecular orbital splitting of the Ni $3d_{3z^2-r^2}$ states inside Ni-Ni dimers. Our calculations reveal that bilayer $La_3Ni_2O_7$ is a strongly correlated magnetic system with robust Ni spin moments. Spin moments on individual Ni sites take on unusually small values because of the electron delocalization over molecular orbitals involving multiple Ni and O sites. We further find that the magnetism of bilayer $La_3Ni_2O_7$ is intimately linked with charge distribution between different Ni and O orbitals. Two distinct regimes are identified in this regard. In one, molecular orbital physics drives the Ni $3d_{x^2-y^2}$ band towards half-filling, which is a well-established condition for unconventional high-temperature superconductivity upon hole doping in cuprates. In the other, the Ni $3d_{x^2-y^2}$ band is quarter-filled favouring spin- and charge-density wave states and Ni-O bond-disproportionation, which is consistent with several recent experimental claims. It is possible that superconductivity in $La_3Ni_2O_7$ occurs as a result of a pressure-induced transition between these two competing regimes. Since none of the low energy phases discovered in this study are metallic, non-stoichiometry would be required for superconductivity to occur.

## I. INTRODUCTION

Unconventional Ni-based superconductivity continues to spark intense scientific interest, now reignited by the recent observation of a superconducting state up to 80 K in pressurized $La_3Ni_2O_7$[1–3]. In distinction from the pioneering infinite-layer (IL) nickelate superconductors[4], $La_3Ni_2O_7$ has a multilayer crystal structure featuring apical oxygen atoms between the $NiO_2$ planes in a multilayer and a high formal Ni oxidation state of 2.5+. The origin of superconductivity in $La_3Ni_2O_7$ is currently highly debated[5–28] and has become obscured even further by this material's recently discovered structural polymorphism[29–31]. The focus of the present study will be on the bilayer variant of $La_3Ni_2O_7$, with double $NiO_2$ layers as shown in Fig. 1, which is particularly interesting due to claims of unique electronic features originating from the molecular orbital (MO) splitting of the Ni $3d_{3z^2-r^2}$ states within the Ni-Ni dimers[8].

Advancing the understanding of $La_3Ni_2O_7$ requires an accurate knowledge of local charge distribution within the $NiO_2$ multilayers, especially in view of potential similarities with the cuprates and/or IL nickelates. In this regard, recent x-ray absorption spectroscopy[32] and electron energy loss spectroscopy[33] experiments on the bilayer $La_3Ni_2O_7$ have confirmed significant presence of oxygen holes in this system. This is what one typically finds in high-valence nickelates[34, 35], such as the cubic $R$NiO$_3$ ($R$=rare-earth atom), as well as in hole doped cuprates. In these materials, the excess positive charge beyond a valence of 2+ for Ni or Cu most likely resides as a hole in the O $2p$ valence states, although strongly hybridized with the transition metal $3d$ states. For the bilayer $La_3Ni_2O_7$, however, a detailed distribution of holes between the three different types of oxygen atoms (planar "$O^p$", inner apical "$O^a$", and outer apical "$O^o$", see Fig. 1) remains to be explored. Since different locations and local symmetries of oxygen hole states are intimately linked with distinct Ni $e_g$ orbital occupancies, such information is key for unveiling the microscopic mechanisms of superconductivity in $La_3Ni_2O_7$ and, more fundamentally, its electronic and magnetic ground state structure. For instance, holes in the planar $O^p$ $2p_x$ and $2p_y$ orbitals are expected to form Zhang-Rice (ZR) singlets with local $x^2-y^2$ symmetries or 3-spin polarons (3SP), which are hallmark features of cuprate-like superconductivity[36–38].

The problem of the local oxygen hole density distribution in bilayer $La_3Ni_2O_7$ has been addressed theoretical by several first principles studies [6, 21, 39–52] that offered many valuable insights but no consensus, nonetheless. A major theoretical challenge is to prop-erly describe strong correlations between the Ni 3$d$ electrons, which necessitates the use of density functional theory (DFT) methods that go beyond the local density approximation (LDA)[53]. Popular methods used to tackle strong correlations include DFT+U[54, 55] and DFT+DMFT (dynamical mean-field theory)[56]. However, both methods involve adjustable parameters (*e. g.*, values of on-site interaction parameters) and multiple choices for implementation schemes (*e. g.*, different double counting corrections and definitions of correlated orbitals), which can have an unphysically strong effect on the results[57]. Moreover, the characteristic MO splitting of the Ni $3d_{3z^2-r^2}$ states within the Ni-Ni dimers presents an additional complication in band theory as it signifies strong *non-local* correlation effects. A proper description of the latter requires the use of elaborate extended schemes in DFT+DMFT[49] and is beyond the applicability of DFT+U[58].

In this paper, motivated to investigate bilayer $La_3Ni_2O_7$ from a relatively less biased *ab initio* standpoint, we choose to employ a hybrid functional approach[59]. It mitigates the self-interaction error by admixing a portion of exact exchange from Hartree–Fock (HF) theory into the DFT exchange-correlation energy. Formulated this way, the hybrid functional approach can involve as few as only one adjustable parameter, which is the mixing ratio of HF and DFT contributions. We should note, however, that within any type of band theory, including hyrid functionals, it is impossible to define a local spin with eigenvalues of $S^2 = S(S+1)$. In contrast, model cluster approaches allow, though at a cost of losing translational symmetry, to avoid this and many other issues, with a recent example coming from the cluster exact diagonalization study by Jiang *et al.* [60]. Appendix A provides further discussion of the dimer problem in band theory.

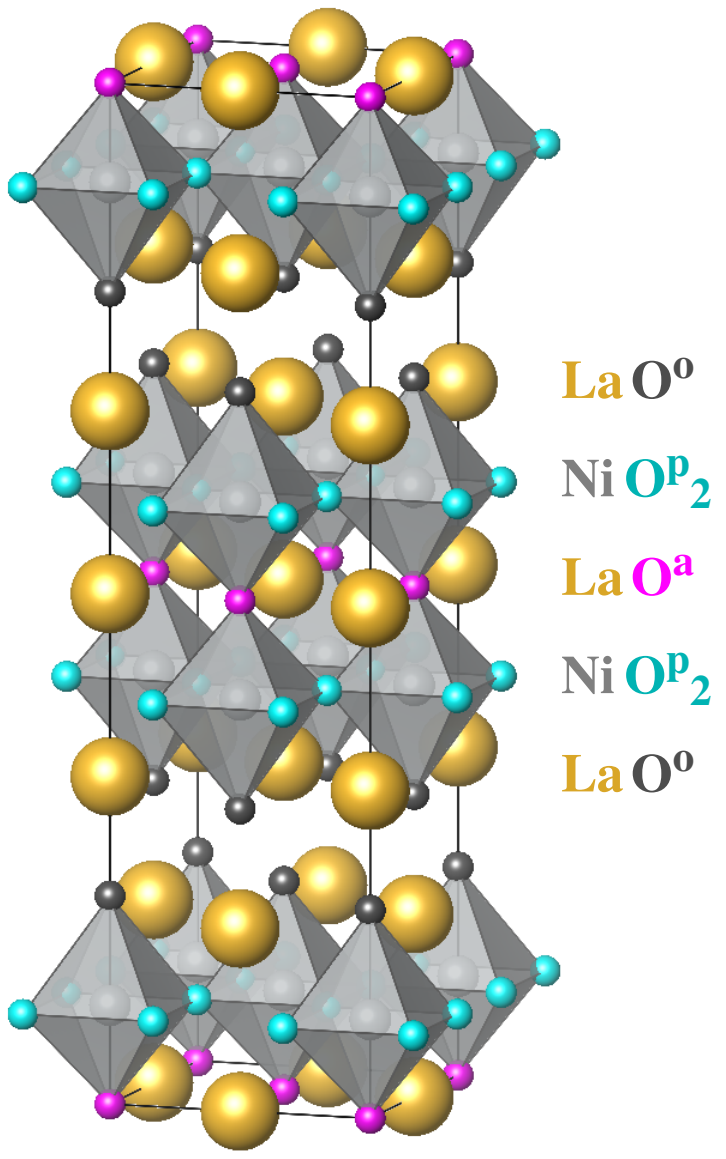

FIG. 1. The $Fmmm$ unit cell of bilayer $La_3Ni_2O_7$, where small balls of different colors indicate positions of the three types of oxygen atoms: planar ("O$^p$"), inner apical ("O$^a$"), and outer apical ("O$^o$").

The most important finding of our hybrid functional study is that the oxygen hole distribution in bilayer $La_3Ni_2O_7$ varies strongly depending on the type of imposed magnetic order. This leads to the formation of a very rich landscape of strongly differing electronic structures for the various possible spin orderings. Some magnetic solutions are found to be metals, others are insulators or small band gap semiconductors. Some form spin- and charge-density-wave (SDW/CDW) states, which can be connected to those observed in transport[61], $^{139}$La-nuclear magnetic resonance (NMR)[62, 63], RIXS[32], neutron powder diffraction (NPD) and muon-spin rotation/relaxation ($\mu$SR) studies[64].

The structure of the paper is as follows. In Sec. II, we will provide computational details of our hybrid functional approach. In Sec. III, we will first review the nonmagnetic electronic structure of bilayer $La_3Ni_2O_7$ and point out the differences in comparison with the previously published LDA results. Then, after discussing possible correlated local configurations in the light of the exact cluster diagonalization findings[60], we will present a selection of magnetically polarized and ordered states of bilayer $La_3Ni_2O_7$ and demonstrate significant variations in oxygen hole distribution patterns among them. We will also use hybrid functional based structural relaxation in order to explore the system's tendencies towards symmetry-breaking bond disproportionation and show that it might be hard to detect it using standard experimental structure characterization techniques. Finally, conclusions will be offered in Sec. IV.

## II. METHODS

Our DFT calculations are performed using the pseudopotential code VASP[65] and the Heyd–Scuseria–Ernzerhof (HSE06) exchange-correlation hybrid functional[66]. When not specified otherwise, the presented electronic structure results are obtained by setting the space group and the lattice constants of bilayer $La_3Ni_2O_7$ to those determined at a temperature of 300 K and an elevated pressure of 29.5 GPa (space group: $Fmmm$, $a = 5.289$, $b = 5.218$, $c = 19.734$ Å) and relaxing atomic positions within the gradient-corrected LDA[67]. We use a $4 \times 4 \times 4$ $\Gamma$-centered $\boldsymbol{k}$-mesh for Brillouin zone integration in $La_3Ni_2O_7$ supercells containing two Ni-Ni dimers. As was confirmed in separate DFT+U calculations, this provides sufficiently well converged total energies and electronic states. The energy cut-off is set to ENCUT=320 eV.

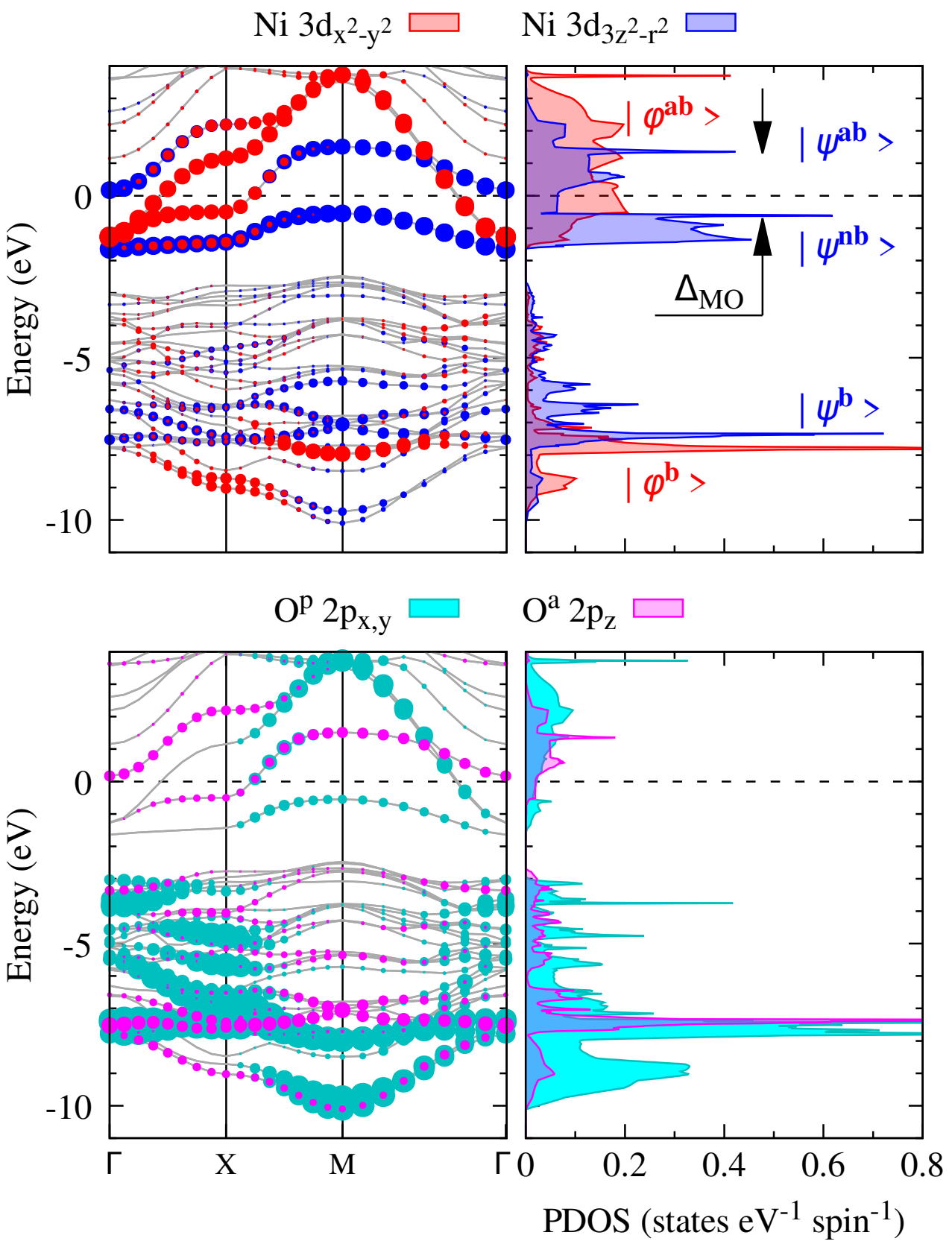

FIG. 2. Non-magnetic electronic structure of bilayer $La_3Ni_2O_7$ calculated with HSE06. The top (bottom) panels display projections onto the Ni $e_g$ (O $2p$) orbitals, highlighting the MO splittings discussed in the main text. The splitting between the molecular $|\psi^{\mathbf{nb}}\rangle$ and $|\psi^{\mathbf{ab}}\rangle$ states, $\Delta_{\mathrm{MO}}$, is indicated by arrows.

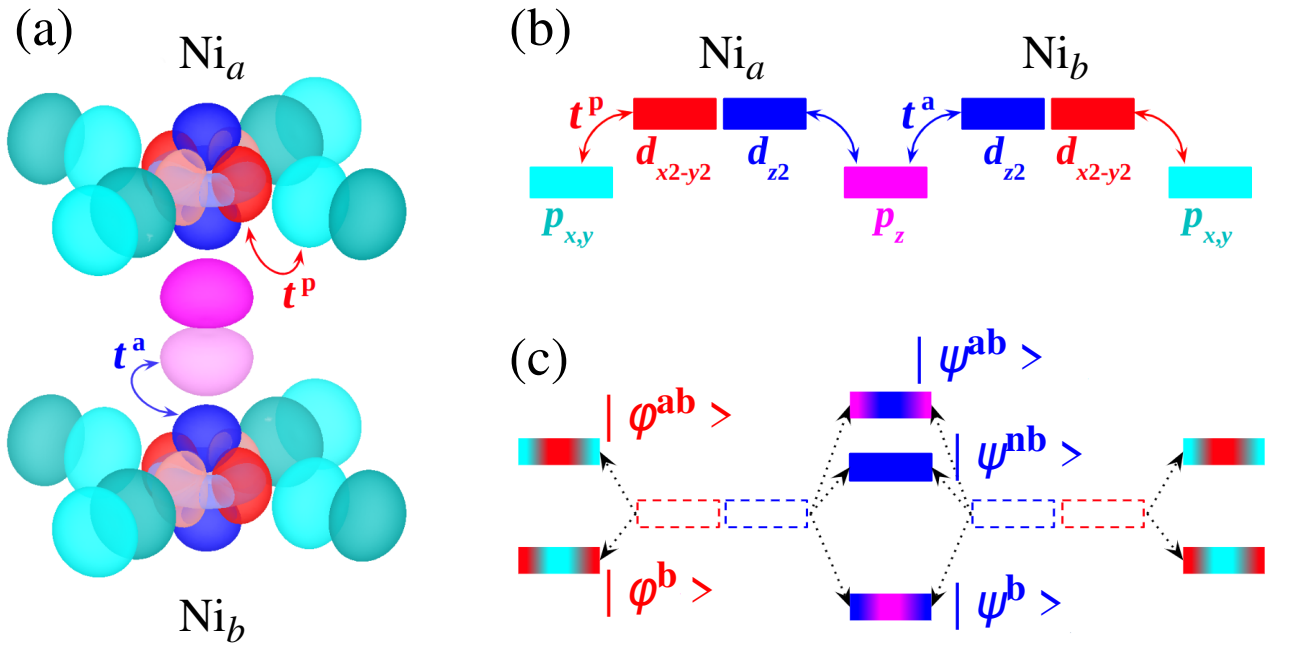

FIG. 3. A molecular orbital description of the $La_3Ni_2O_7$ electronic structure. (a) Ni $e_g$ and relevant O $2p$ orbitals in a Ni-Ni dimer together with the $t^{\mathrm{p}}$ and $t^{\mathrm{a}}$ hopping integrals. (b) Energy levels of the Ni $e_g$ orbitals on Ni$_a$ and Ni$_b$ as well as of the inner apical O$^{\mathrm{a}}$ $2p_z$ and planar O$^{\mathrm{p}}$ $2p_{x,y}$ orbitals in the Ni-Ni dimer. (c) Energy levels of the MO states resulting from Ni-O hybridization.

## III. RESULTS AND DISCUSSION

### A. Non-magnetic electronic structure from HSE06: stabilization of the molecular orbital splitting

Let us first review key atomic orbitals and hybridization interactions that shape the electronic structure of bilayer $La_3Ni_2O_7$. These are well exposed in Figure 2, showing orbital projected electronic states of $La_3Ni_2O_7$ from a non-magnetic HSE06 calculation. This HSE06 result (also previously reported in Ref. 50) is qualitatively similar with the LDA one[1], apart from the increased Ni-O states' bandwidth (from about 11 eV to 14 eV). Within both approximations, electronic structure near the Fermi level is dominated by Ni $e_g$ orbitals. The Ni $3d_{x^2-y^2}$ orbitals are strongly hybridized with the planar oxygen O$^{\mathrm{p}}$ $2p_x$ and $2p_y$ orbitals lying in the same $NiO_2$ plane via the hopping integral $t_{\mathrm{p}}$, as shown in Figs. 3 (a, b). As a result, two MO *band* states (per plane) are formed:

$$(\mathbf{bonding}) \quad |\phi^{\mathbf{b}}(\boldsymbol{k})\rangle = \alpha(\boldsymbol{k})|d_{x^2-y^2}\rangle + \beta(\boldsymbol{k})\left(\sum_{i=1}^{4}\eta_i(\boldsymbol{k})|p_\sigma\rangle_i\right)$$

and

$$(\mathbf{anti\text{-}bonding}) \quad |\phi^{\mathbf{ab}}(\boldsymbol{k})\rangle = \beta(\boldsymbol{k})|d_{x^2-y^2}\rangle - \alpha(\boldsymbol{k})\left(\sum_{i=1}^{4}\eta_i(\boldsymbol{k})|p_\sigma\rangle_i\right),$$

where the subscripts $i = 1,2,3,4$ run over the four corners of the oxygen square around a central Ni $3d_{x^2-y^2}$ orbital, $|p_\sigma\rangle_i$ are the $p_x$ and $p_y$ states centered at those corners that are $\sigma$-bonded to the central Ni $3d_{x^2-y^2}$ orbital, and $\eta_i(\boldsymbol{k})$ are the symmetry respecting phases of individual oxygen atomic orbitals in the arising oxygen MO. This splitting is illustrated in Fig. 3 (c). The anti-bonding MO band crosses the Fermi level in a way similar to what is observed in the cuprates, apart from the fact that here this band is quarter-filled. Note that the normalizing coefficients $\alpha(\boldsymbol{k})$ and $\beta(\boldsymbol{k})$ are strongly $\boldsymbol{k}$-vector dependent.

In their turn, the $3d_{3z^2-r^2}$ orbitals of the two Ni atoms in a dimer [labeled as Ni$_a$ and Ni$_b$ in Figs. 3 (a, b)] each hybridize nearly as strongly with the $2p_z$ orbital of the central apical oxygen atom O$^{\mathrm{a}}$ via the hopping integral $t_{\mathrm{a}}$. This gives rise to the following three MO states:

$$(\mathbf{bonding}) \quad |\psi^{\mathbf{b}}\rangle = \gamma(|d_{z^2}\rangle_a - |d_{z^2}\rangle_b) + \delta|p_z\rangle,$$

at around $-8$ eV,

$$(\mathbf{non\text{-}bonding}) \quad |\psi^{\mathbf{nb}}\rangle = (|d_{z^2}\rangle_a + |d_{z^2}\rangle_b)/\sqrt{2},$$

at $-1$ eV, and

$$(\mathbf{anti\text{-}bonding}) \quad |\psi^{\mathbf{ab}}\rangle = \delta(|d_{z^2}\rangle_a - |d_{z^2}\rangle_b) - \gamma|p_z\rangle,$$

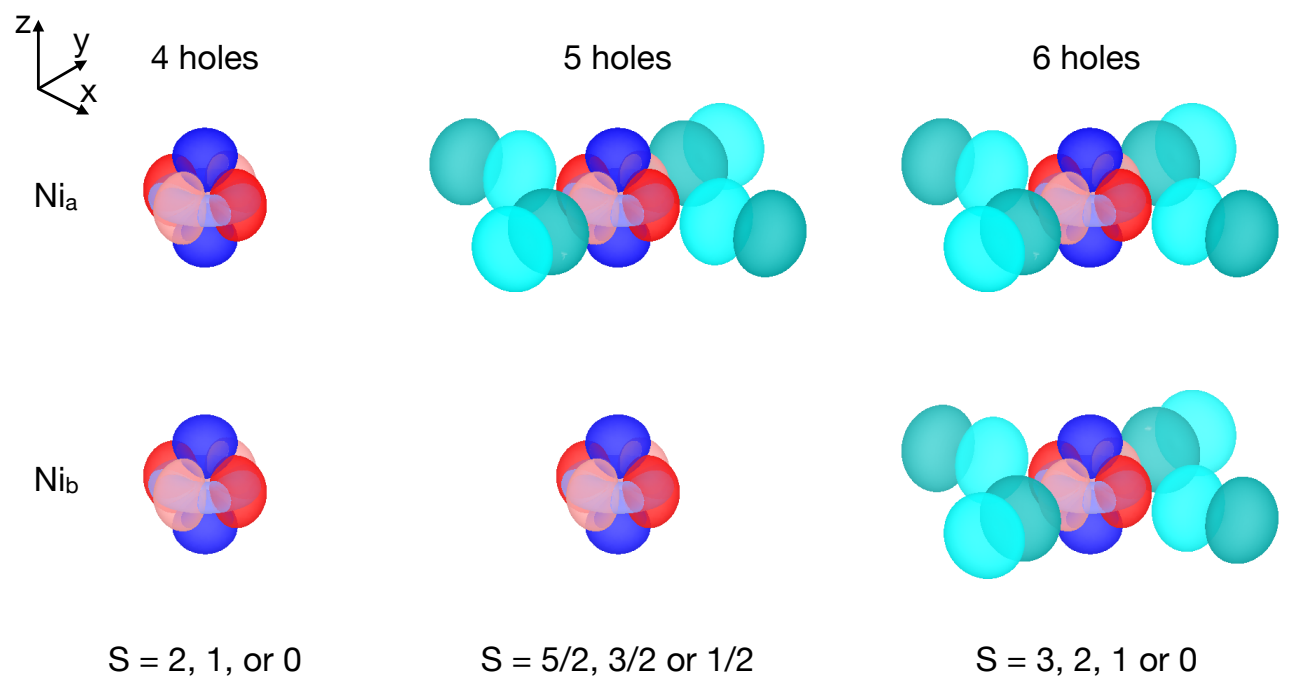

FIG. 4. An artist's concept of the lowest energy hole distributions in Ni dimers. Possible spin states are indicated at the bottom. Left dimer has 4 holes occupying $3z^2-r^2$ and $x^2-y^2$ Ni $3d$ orbitals. Middle dimer has one extra hole in a linear combination of Oxygen $p_x$ and $p_y$ orbitals centred about one Ni with $x^2-y^2$ symmetry. Both Ni ions have ligand holes associated with them as shown on the right for a 6 hole configuration.

at 1 eV. The "$a$" and "$b$" subscripts here refer to the two Ni atoms in the dimer. The weak dispersion of these states in the $xy$ plane is due to Ni $3d_{3z^2-r^2}$ hybridizing with the O$^p$ $2p_{x,y}$ combinations of the $x^2+y^2$ symmetry. Note that, unlike $\alpha(\boldsymbol{k})$ and $\beta(\boldsymbol{k})$, the normalizing coefficients $\gamma$ and $\delta$ are essentially independent of the $\boldsymbol{k}$-vector. In agreement with Ref. 50, we find that HSE06 corrects the underestimation of the bilayer coupling in LDA by increasing the energy separation between states $|\psi^{\mathbf{nb}}\rangle$ and $|\psi^{\mathbf{ab}}\rangle$, $\Delta_{\mathrm{MO}}$, from 1.5 to 2 eV. We would also like to note that the non-bonding state $|\psi^{\mathbf{nb}}\rangle$ has no O$^a$ $2p_z$ component because of an opposite parity between $|d_{z^2}\rangle_a+|d_{z^2}\rangle_b$ and $|p_z\rangle$.

Before concluding this section, we point out that oxygen orbitals other than O$^p$ $2p_{x,y}$ and O$^a$ $2p_z$ do not appear to play any significant role in the physics of bilayer $La_3Ni_2O_7$, including the $2p$ orbitals of the outer apical oxygen O$^o$.

### B. Hole distribution configurations for a Ni-Ni dimer

While non-magnetic LDA or HSE06 calculations are helpful in identifying key orbitals and hybridization interactions, local magnetism, which is absent in those calculations, may affect orbital occupations dramatically. In this regard, let us review the magnetic states of the hole distribution configurations within dimers containing four, five, and six holes (Fig. 4) that are found to be the *lowest in energy* in the exact diagonalization calculations of Jiang *et al.*[60]. They all have no hole in O$^a$ $2p_z$. We continue to label Ni atoms within the same dimer as "$a$" and "$b$". We will also consider two nearest-neighbour dimers and label them as "$A$" and "$B$".

There are five holes per dimer in bilayer $La_3Ni_2O_7$ on average. They can get ordered as $A(a^3b^2)B(a^2b^3)$, which corresponds to a quarter filled band in both $NiO_2$ planes, but 5 holes on each dimer staggered on $a$ and $b$. Also possible is configuration $A(a^3b^3)B(a^2b^2)$, corresponding to alternation of 6 and 4 hole dimers, which can be achieved by transferring one hole from $A$ to $B$, *i. e.*, a minimal charge fluctuation.

Let us now look at the spins. All the three-hole states involve a ZR singlet whose exchange interaction dominates strongly over the Hund's rule exchange, leaving an $s=1/2$ in the $d_{3z^2-r^2}$ orbital. All the two-hole cases are $d^8$ and with a spin of 1 (Hund's rule exchange). Therefore, the spin on dimer $A(a^3b^2)$ is $a_{s=1/2}b_{s=1}$, and on $B(a^2b^3)$ it is $a_{s=1}b_{s=1/2}$. The intradimer exchange involving the $3z^2-r^2$ orbitals on each $a$ and $b$ sites is antiferromagnetic and a result of superexchange between the two Ni via O$^a$ of about 50 meV. This favours the dimer spin-1/2 state. Band theory, in which spin is not a good quantum number, would see this as $A(a_{s_z=+1/2}, b_{s_z=-1})$ and $B(a_{s_z=+1}, b_{s_z=-1/2})$ for a ferromagnetic in-plane coupling, and reversed signs for $B$ for an antiferromagnetic coupling.

Now, we can also have the case of the alternating 6-hole - 4-hole dimers, $A(a^3, b^3)B(a^2, b^2)$, in which the ZR singlets would have $s=0$, leaving the $3z^2-r^2$ orbital with spin 1/2. The exchange interactions in this configuration would prefer a spin zero state for dimer $A$ and spin 0 or 1 state for dimer $B$, which are closely spaced. However, DFT, which has only $S_z$ projections, will see $A(a_{s_z=+1/2}, b_{s_z=-1/2})$ and $B(a_{s_z=+1}, b_{s_z=-1})$, or signs reversed for an antiferromagnetic in-plane coupling. This discussion illustrates that the results of exact diagonalization go generally beyond single particle band theories, unless there is a strong long-range magnetic order, in-plane and between planes.

For the less energetically favourable case of the fifth hole being located in the O$^a$ $2p_z$ orbital [see Fig. 7 (b)], the inter-plane coupling would be ferromagnetic because this O hole and the two $3z^2-r^2$ orbitals would want to form a 3-spin polaron [38] with the Ni spins parallel. As we will demonstrate later, an interesting feature of this configuration is that it preserves the MO splitting between the $|\psi^{\mathbf{nb}}\rangle$ and $|\psi^{\mathbf{ab}}\rangle$ states. However, mirroring the exact diagonalization findings, hybrid functional calculations find this configuration at a relatively higher energy, even with no quantum spin fluctuations taken into account.

### C. Magnetically polarized HSE06 solutions and their distinct oxygen hole distribution patterns

With these considerations in mind, let us now switch the magnetism on in hybrid calculations. We find that a magnetically polarized state with a finite magnetic moment on Ni is by more than 0.75 eV per Ni lower in energy compared with a non-magnetic one. Most of this per Ni energy lowering is a result of the Hund's exchange between the Ni $3d_{3z^2-r^2}$ and $3d_{x^2-y^2}$ orbitals with spin parallel electron occupations and is indeed close to the value of 0.65 eV established for $Ni^{2+}$ in the $d^8$ configuration. Our finding is in agreement with RIXS[32] and muon spin relaxation[68] experiments detecting the presence of localized magnetic moments in $La_3Ni_2O_7$.

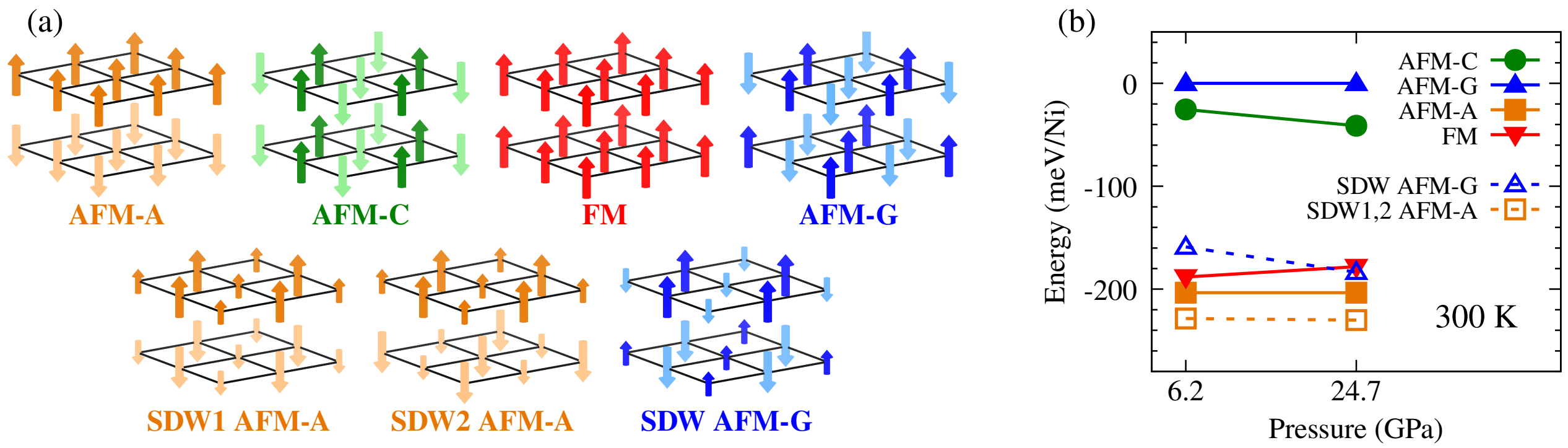

FIG. 5. (a) Magnetic configurations of bilayer $La_3Ni_2O_7$ considered in the main text and (b) their relative total energies per Ni calculated in HSE06 using low- and high-pressure structural parameters. In (a), arrows represent the orientations of Ni spin magnetic moments at different sites of the bilayer square lattice.

We will now consider a selection of magnetically ordered configurations of $La_3Ni_2O_7$ that will serve as computationally accessible proxies to its true, strongly fluctuating, magnetic ground state. The seven Ni spin configurations in focus are displayed in Fig. 5 (a), while their relative energies at different pressure values can be found in Fig. 5 (b) and their energy gaps and Ni magnetic moments in Table I. These configurations include a ferromagnetic (FM) one, three antiferomagnetic ones (AFM-G, AFM-C, AFM-A) and three spin-density-wave ones (SDW AFM-G, SDW1 AFM-A, SDW2 AFM-A). The AFM-A and FM configurations, both characterized by an FM in-plane spin ordering, are among the lowest energy states in HSE06, which is in qualitative agreement with DFT+U calculations of Geisler *et al.*[47]. As one of the most important results of this work, we find that some of the uniform phases display spin- and charge-density-wave instabilities. In the three discovered SDW/CDW phases, which include SDW AFM-G and two nearly degenerate AFM-A phases: SDW1 AFM-A, and SDW2 AFM-A, the system gains energy by having Ni magnetic moments on different lattice sites disproportionate in magnitude, as indicated in Fig. 5 (a) by arrows of different sizes.

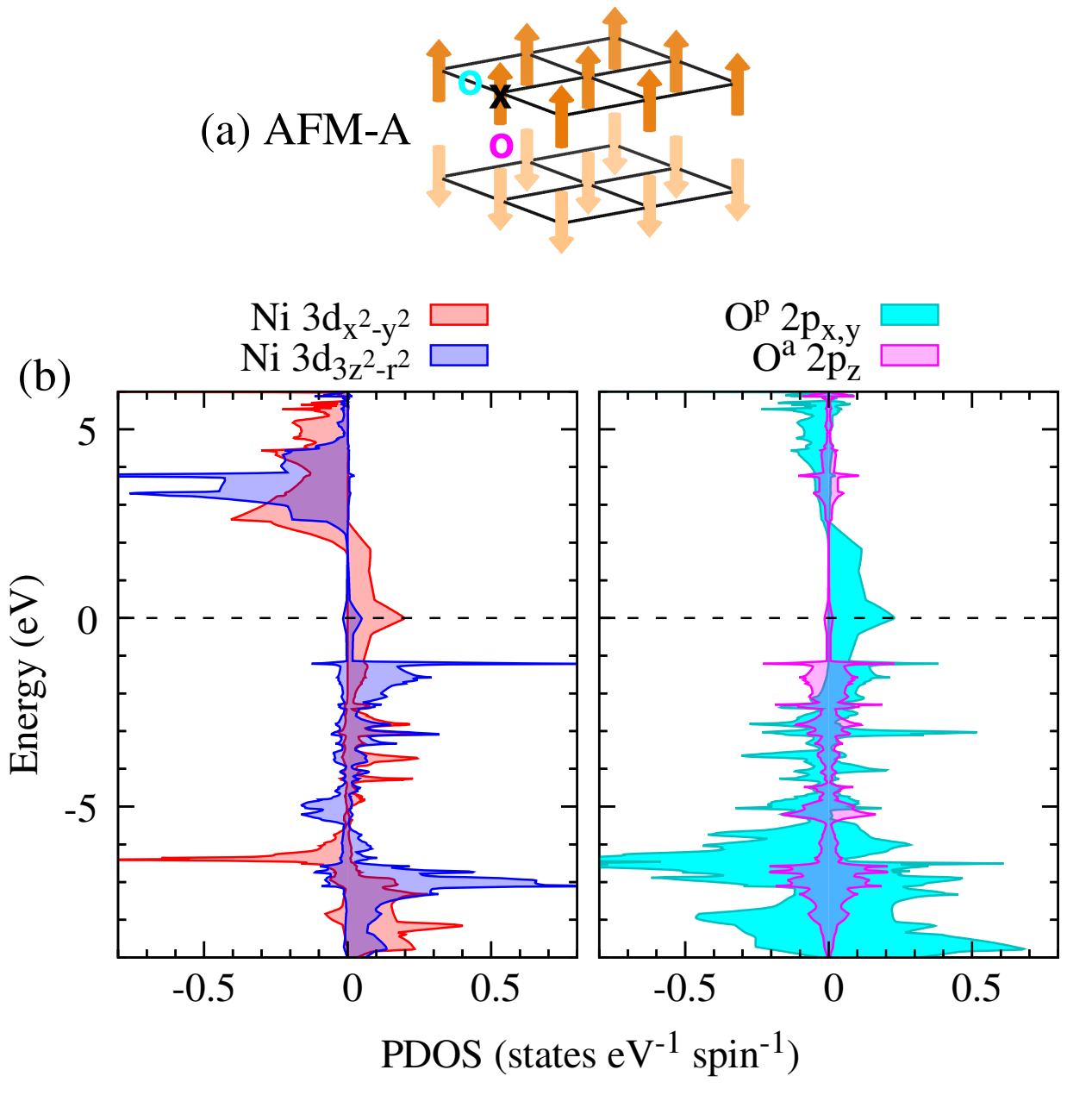

FIG. 6. HSE06 electronic structure of the uniform AFM-A phase in bilayer $La_3Ni_2O_7$: (a) the AFM-A magnetic configuration and (b) the spin-resolved projected densities of states (PDOSs) for the Ni $e_g$ orbitals (left panel) and selected O $2p$ orbitals (right panel) with multiplicity factors taken into account. Here in (a) and also in the subsequent similar figures, the "x" and "o" symbols indicate the Ni or O atoms for which the PDOSs in (b) were generated for.

**AFM-A.** Let us first focus on the uniform moment solutions. The lowest energy uniform solution is AFM-A, with ferromagnetic planes coupled antiferromagnetically. The ferromagnetic in-plane coupling is the result of the antiferomagnetic coupling of an $O^p$ $2p_{x,y}$ hole in the bridging position between two Ni's to their parallel spins of 1 each. In cuprates, the three spins coupled in such a fashion are described in terms a 3-spin polaron with the lowest energy state of total spin 1/2[38]. Here, it would be a 5-spin polaron because of the additional spins of the $3z^2 - r^2$ holes, and the lowest energy state would have total spin 3/2. With oxygen holes so involved in 5-spin polarons in the planes, apical oxygens $O^a$ are left with closed $2p$-shells, providing the antiferromagnetic exchange between the Ni spins in a dimer. This is consistent with the oxygen PDOS in Fig. 6, where $O^a$ $2p_z$ is all but fully occupied with equal amounts of spin up and spin down. We should note also that the hole den-sity of states just above $E_{\rm F}$ has an almost equal amount of Ni $3d_{x^2-y^2}$ and $O^p$ $2p_{x,y}$ orbital characters, consistent with a very strong hybridization between them and consistent with the 3SP picture. In DFT, the 3SP like state has Ni $z$-axis projected magnetic moments that are equal and sharing the $S_z$ projection of $S = 3/2$, which results in a local $z$-projected moment of about 1.5 $\mu_{\rm B}$. As one can see in Table I, this value is reduced to 1.32 $\mu_{\rm B}$ by the covalent hybridization with oxygen.

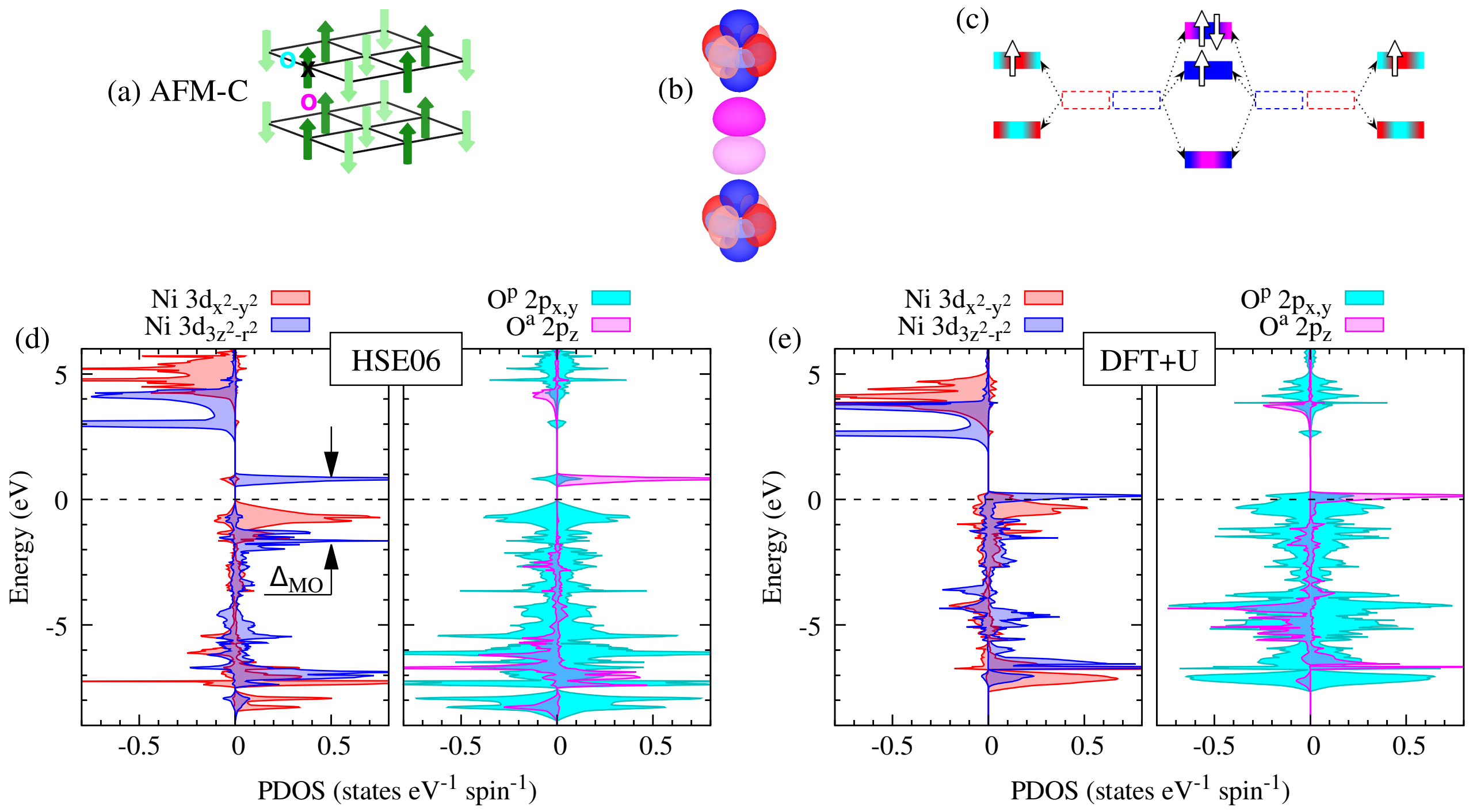

FIG. 7. Electronic structure of the AFM-C magnetic phase in bilayer $La_3Ni_2O_7$: (a) the AFM-C magnetic configuration, (b) the Ni-Ni dimer with a hole in $O^a$ $2p_z$ (c) schematic distribution of the five holes (up and down arrows) between the MO states in a Ni-Ni dimer, (d) HSE06 spin-resolved PDOS for the $e_g$ orbitals of one Ni in the dimer (left panel) and for the apical O $2p_z$ and planar O $2p_{x,y}$ orbitals (right panel), with multiplicity factors taken into account, (e) same as (d) but obtained within DFT+U. In (d), the splitting between the molecular states $|\psi^{\mathbf{nb}}\rangle$ and $|\psi^{\mathbf{ab}}\rangle$, $\Delta_{\rm MO}$, is indicated by arrows.

TABLE I. HSE06 energy gaps and Ni magnetic moments of the seven considered magnetic configurations in bilayer $La_3Ni_2O_7$.

| | Gap (eV) | $\mu$ ($\mu_{\rm B}$) |
|---|---|---|
| FM | 0.00 | 1.40 |
| AFM-C | 0.52 | 1.25 |
| AFM-G | 0.00 | 1.09 |
| AFM-A | 0.00 | 1.32 |
| SDW AFM-G | 1.22 | 1.50/0.41 |
| SDW1 AFM-A | 0.77 | 1.47/1.13 |
| SDW2 AFM-A | 0.71 | 1.45/1.14 |

**AFM-C.** Now, let us compare the AFM-A configuration with the PDOS of the less energetically favourable AFM-C phase, shown in Fig. 7 (d), where the planes are in a G-type antiferromagnetic arrangement with inter-plane Ni dimer coupling being ferromagnetic. The electronic structure has changed dramatically! The material is now a 1 eV gap semiconductor with the oxygen hole density just above $E_{\rm F}$ occupying the inter-plane apical oxygen $O^a$. (We note that DFT+U yields a metallic solution [Fig. 7 (e)], which is discussed in Appendix B.) These empty $O^a$ $2p_z$ states are involved in the $|\psi^{\mathbf{nb}}\rangle$ molecular orbitals discussed earlier and form an extremely narrow band because of the small inter-dimer electron hopping. It is now the $O^a$ holes in between the planes that provide the ferromagnetic coupling between the Ni $3d_{3z^2-r^2}$ holes and thus reposition the 5-spin polarons from the planes to the dimers. In band theory, the 5-spin polarons are mimicked by spin polarization of the $|\psi^{\mathbf{nb}}\rangle$ molecular orbitals, as illustrated in Fig. 7 (c). Similarly to the AFM-A case, covalency effects are found here to reduce the Ni $z$-axis projected magnetic moments from 1.5 $\mu_{\rm B}$, expected for the net $S_z = 3/2$ shared between the two Ni sites in a dimer, to 1.25 $\mu_{\rm B}$ (Table I). An intriguing feature of this particular oxygen hole density distribution is that the anti-bonding $x^2-y^2$-symmetry MO band $|\phi^{\mathbf{ab}}(\boldsymbol{k})\rangle$ is now exactly half-filled, which closely parallels the situation in cuprates. However, due to the relatively higher energy of the AFM-C phase, this electronic configuration is unlikely to be realized in bilayer $La_3Ni_2O_7$.

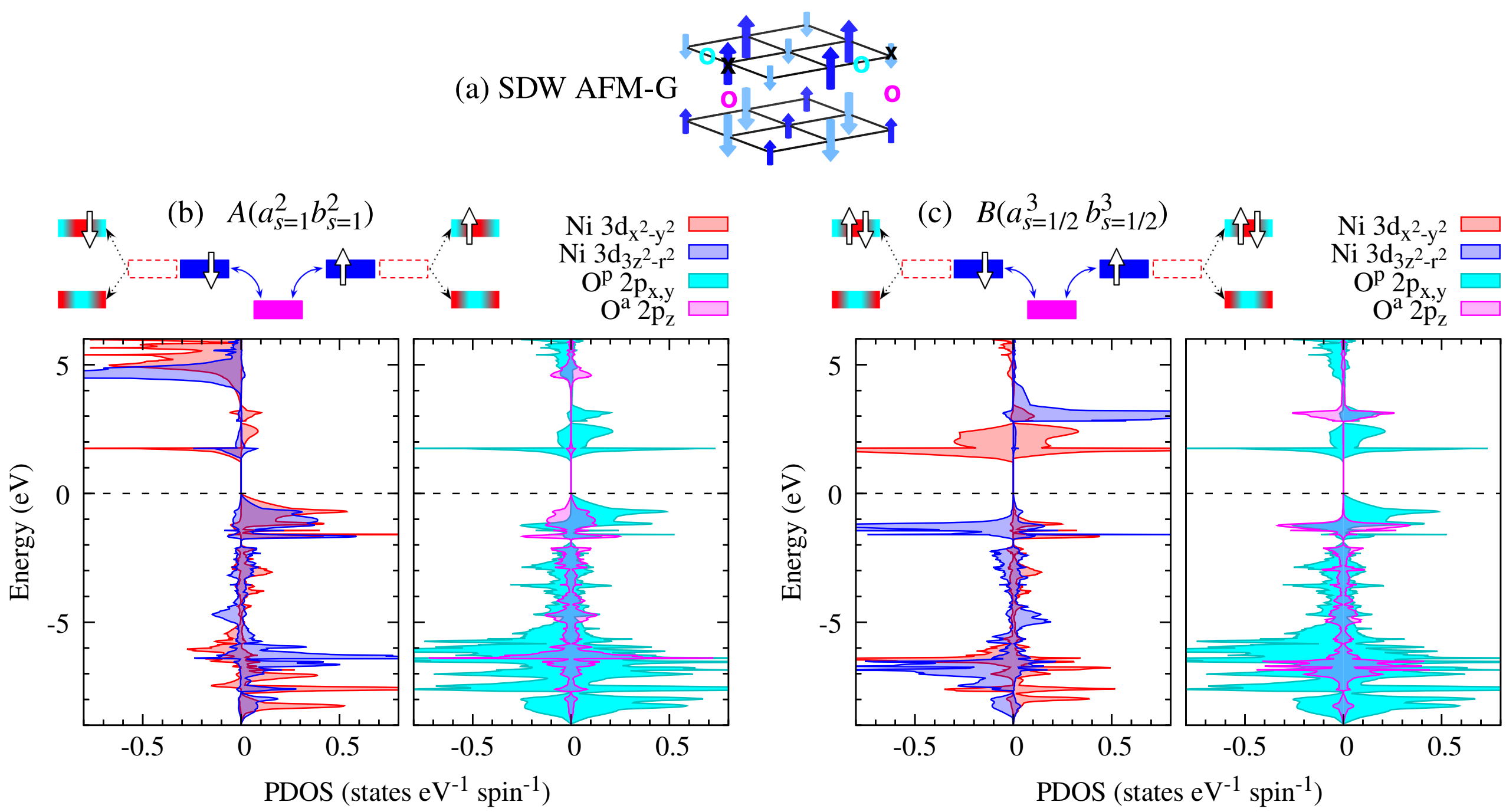

FIG. 8. HSE06 electronic structure of the SDW AFM-G phase in bilayer $La_3Ni_2O_7$: (a) the SDW AFM-G magnetic configuration, with a checkerboard order of large and small Ni spin dimers, (b,c) schematic distributions of holes and PDOS for the (b) large and (c) small spin Ni dimers. In (a), the three colored "x" and "o" symbols on the left pertain to the PDOSs shown in (b) and those three on the right to the PDOSs in (c).

**SDW AFM-G.** We now look at the non-uniform moment SDW cases. These phases are generated by disproportionated magnetic moments from the AFM-G and AFM-A phases. Their PDOSs are shown in Figs. 8-10. These SDW phases are all at lower energies (by about 30 or 160 meV per Ni) than the parent phases, which is a considerably strong stabilization. They each represent a checkerboard pattern of large and small moments in each of the two planes. The AFM-G SDW phase has the same type of antiferromagnetic ordering both in the planes and between the planes with an in-phase checkerboard pattern of small and large moments. The PDOS in Fig. 8 shows us a very large low-energy extra Ni $3d_{x^2-y^2}$ hole density with a strongly diminished in-plane O$^p$ $2p_{x,y}$ occupation. It also shows us a strong CDW accompanying the SDW. Going back to our notation of the charge ordering as between two dimers labelled by $A$ and $B$ and within each dimer between two Ni's labelled as $a$ and $b$, we would label this configuration as $A(a^2_{s=1}, b^2_{s=1})B(a^3_{s=1/2}, b^3_{s=1/2})$, *i. e.*, the four hole − six hole case. Interestingly, the large spin here has the smaller hole density, which means that to add holes but lower the spin we must be forming $s = 0$ states between pairs of spins. This can be done with either ZR singlets or doubly occupied Ni $3d_{x^2-y^2}$ states. Looking at the low-spin and large hole density dimer $B$ [Fig. 8 (c)] just above $E_F$, it seems to be the $d^7$ state, with two holes in the Ni $3d_{x^2-y^2}$ orbitals and one in the Ni $3d_{3z^2-r^2}$ orbital, that causes the low spin, rather than the ZR singlet plus a Ni $3d_{3z^2-r^2}$ hole. We know from the cluster calculations that the $d^7$ $s = 1/2$ state is close in energy to the ZRS plus a Ni $3d_{3z^2-r^2}$ hole state also of spin 1/2.

**SDW1 and SDW2 AFM-A.** We now turn to the AFM-A SDW phases, of which there are two, with nearly the same energies (see Figs. 9 and 10). The AFM-A phase has ferromagnetic planes coupled antiferromagnetically. This is also the case in the two SDW phases, with also ferromagnetic planes coupled antiferromagnetically but with each having a checkerboard pattern of alternating high- and low-spin Ni magnetic moments. In SDW1, the spin size alternating patterns of the two planes are in phase, while in SDW2 they are in antiphase. With regard to the hole occupations, the first one has the same four hole − six hole ordering as in SDW AFM-G, and the second one has the five − five, *i. e.*, $A(a^2_{s=1}, b^3_{s=1/2})B(a^3_{s=1/2}, b^2_{s=1})$, hole ordering. As one can see in Figs. 9 and 10, their PDOSs are rather similar, indicating that super exchange via the inter-planar O$^a$ does not affect charge distribution much. Indeed, this super exchange interaction is only about 50 meV according to the cluster calculations while the ZRS exchange interaction between the ligand hole ($\underline{L}$) on $O^p$ and the Ni $3d_{x^2-y^2}$ hole is very large between 1 and 2 eV. We also see from the PDOSs that the large-spin dimer again has little $O^p$ $2p_{x,y}$ hole density, while the low-spin one has a lot but also a lot of Ni $3d_{x^2-y^2}$ low-energy hole density, similar to what we see in the AFM-A phase, which we designated as the result of a 3SP involving two Ni's and the in-between $O^p$ rather than a ZR singlet, with now significant $p_{x,y}$ and $d_{x^2-y^2}$ densities of the same spin at low hole energies.

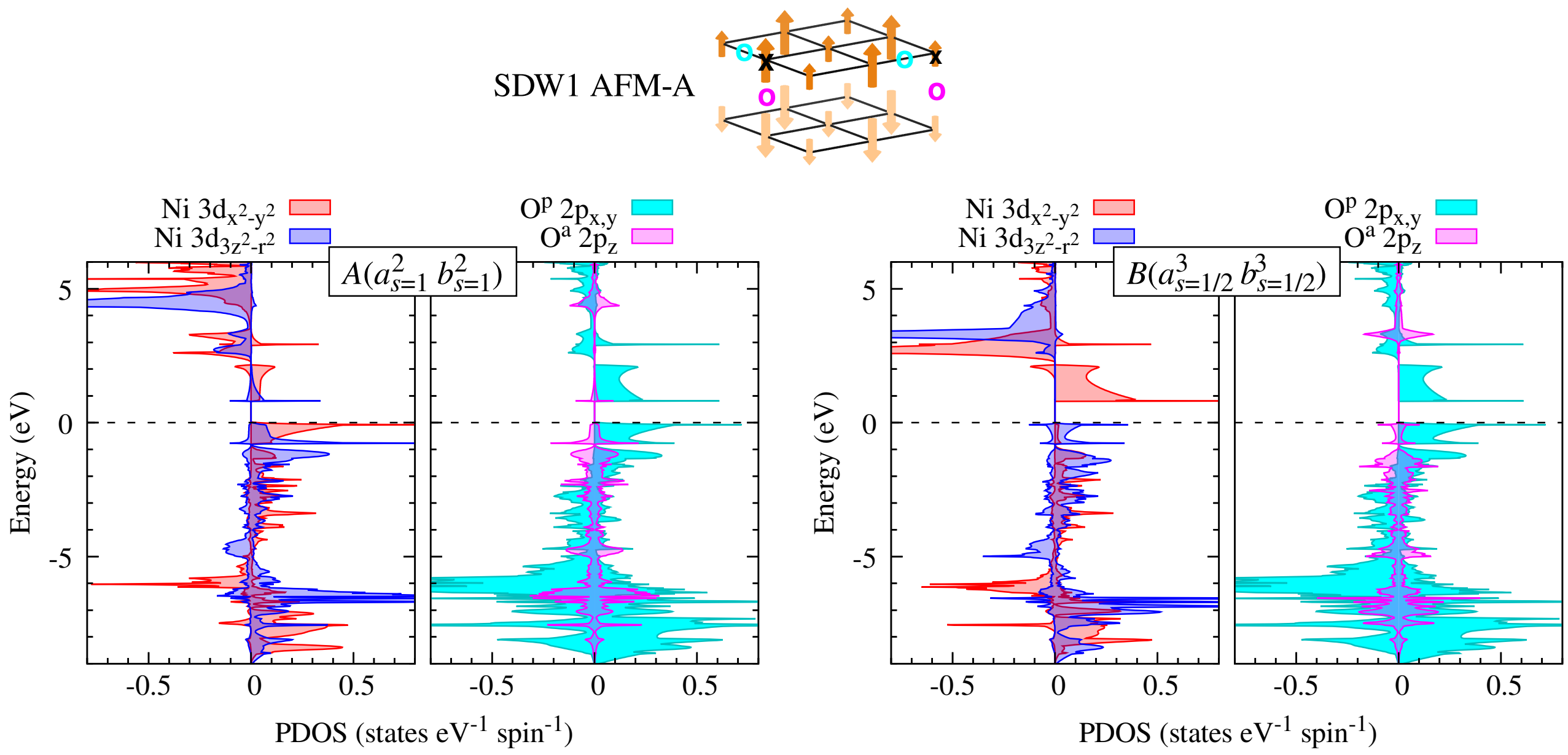

FIG. 9. HSE06 electronic structure of the SDW1 AFM-A phase in bilayer $La_3Ni_2O_7$: (a) the SDW1 AFM-A magnetic configuration, with a checkerboard order of large and small Ni spin dimers, (b,c) PDOSs for the (b) large and (c) small spin Ni dimers.

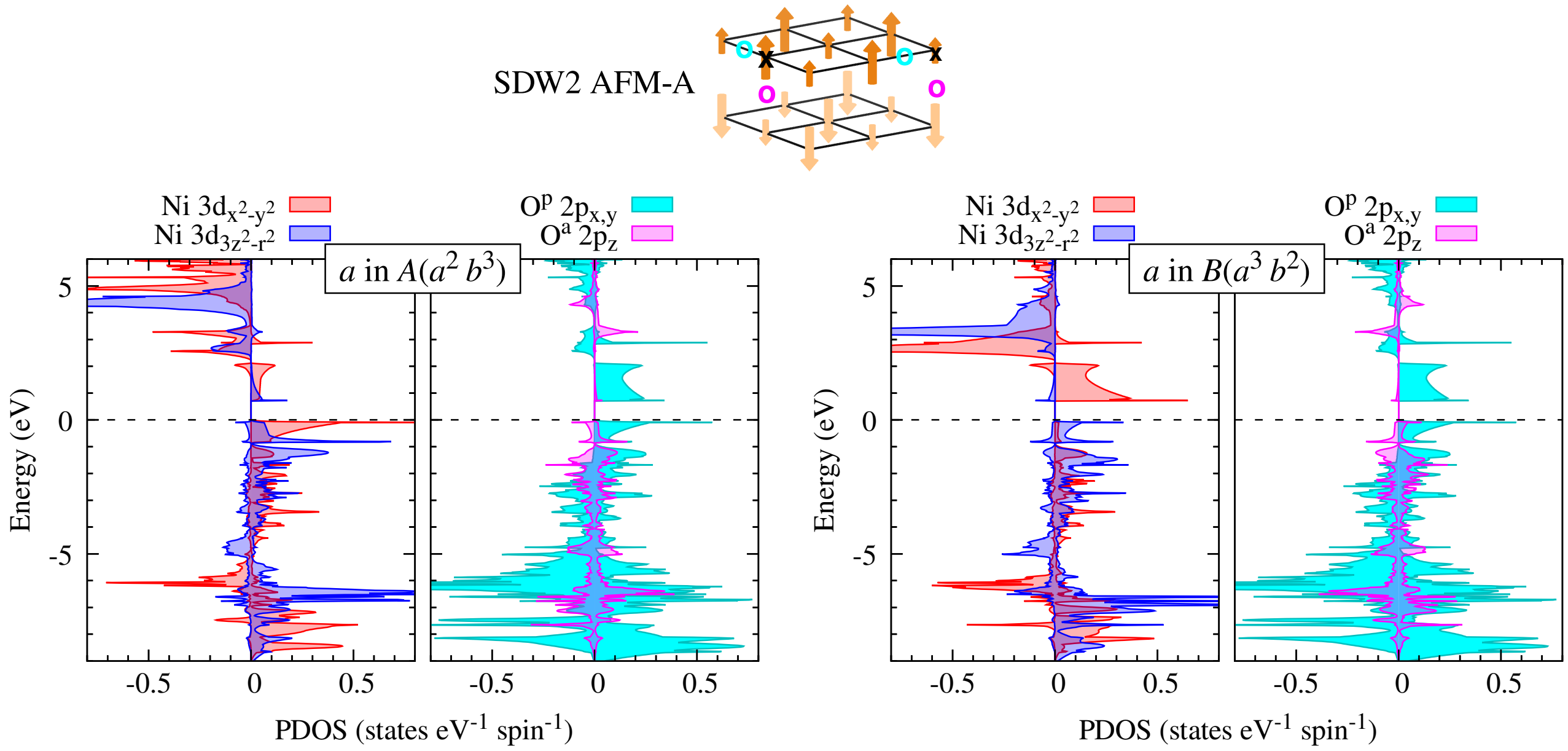

FIG. 10. HSE06 electronic structure of the SDW2 AFM-A phase in bilayer $La_3Ni_2O_7$: (a) the SDW2 AFM-A magnetic configuration, with a checkerboard order of large and small Ni spin sites in each plane and with a phase shift between the planes, (b,c) PDOSs for the (b) large and (c) small spin Ni sites.

Making now a connection with the real picture, one should keep in mind that all band-structure applications of DFT, including the hybrid functional approach, produce single-Slater-determinant single-particle states restricted by translational symmetry, which consider $z$-projections of spins to be good quantum numbers and not the spins themselves. This creates a real problem for states that are locally antiferromagnetically coupled. For example, spin-1/2 antiferromagnetically coupled states form singlets and triplets, with the singlets at the energy of $-\frac{3}{4}J$ and the triplets at $+\frac{1}{4}J$. However, if we assume Ising spins and consider only the $S_z$ component, the energies are $-\frac{1}{4}J$ and $+\frac{1}{4}J$. So, the stabilization of singlet-like states with zero net spin is $-\frac{1}{2}J$ lower than that generated by assuming Ising spins. This means that all the states that potentially involve local singlets will in reality be considerably lower in energy than the $S_z = 0$ states generated by band-structure methods. (We note that the singlet states of our concern here should be distinguished from the $S = S_z = 0$ states of two identical orbitals being occupied by a spin-up and a spin-down electron because of the Pauli exclusion principle and not because of exchange.) The underestimation of the singlet stabilization is especially significant if the number of neighbours a central spin is coupled to is small. Indeed, if it is smaller than three, then the singlet solution is generally much more stable than the $S_z$ only spin ordered states. For bilayer $La_3Ni_2O_7$, this especially strongly influences the inter-plane or intra-cluster exchange between the Ni $3d_{3z^2-r^2}$ orbitals generated by the coupling to an $O^a$ $2p_z$ hole, *i. e.*, the 3SP case. These states involve very strong quantum fluctuations resulting in a net spin that is delocalized over the cluster, which complicates the effective inter-cluster interactions and the global magnetic structure. In general, all states involving an O hole either in-plane or between planes that is strongly antiferromagnetically coupled to its nearest-neighbour Ni spins will in reality have considerably lower energies than what is predicted by band-structure methods.

Concluding this part, we see a strong relationship between the charge density and the spin density and the type of coupling, be it ferromagnetic or antiferromagnetic, which strongly complicates the electronic and magnetic structure of the lowest-energy states. Our conclusion is that the two nearly degenerate AFM-A SDW phases are the lowest-energy ones. It is interesting to note that in these lowest-energy phases the in-plane spin correlation is ferromagnetic. Of course, it is quite possible that for a larger supercell we could have a longer wave-length SDW/CDW, involving, for example, ferromagnetic pairs coupled antiferromagnetically.

### D. Bond disproportionation in the SDW/CDW states

All of the considered SDW/CDW states are found to be further stabilized by Ni-O bond disproportionation, whereby alternating large and small oxygen squares are formed around the $d^8$ and $d^8\underline{L}_{x^2-y^2}$ Ni sites, respectively. This also further favours the ZRS like states in the small squares relative to the 3SP like description. Table II summarizes the relative total energies and Ni-O bond lengths of the AFM-C, AFM-G, SDW AFM-G, and SDW1 AFM-A states discussed above. These results are obtained by performing full relaxation of atomic positions within HSE06. Full structural information on these four fully optimized structures can be found in Appendix C. Comparing Table II and Fig. 5 (b), one can see that allowing for bond disproportionation increases the energy separation between the uniform Néel and SDW/CDW states by about 100 meV per Ni.

Curiously, we find that even though the Ni-O bond disproportionation is quite significant and also reduces the structural symmetry of bilayer $La_3Ni_2O_7$ (for instance, from $Amam$ to $Amm2$ at $T = 300$ K and $P = 1.6$ GPa), this could be hard to detect with standard x-ray diffractometry. Figure 11 shows simulated diffraction patterns of the fully relaxed bilayer $La_3Ni_2O_7$ in the AFM-C and SDW AFM-G states at $T = 300$ K and $P = 1.6$ GPa. The simulations have been done using the VESTA code[69]. The differences between the two diffractograms are barely detectable and might be beyond the experimental resolution. We therefore urge experimental groups to reexamine results of structural characterization of this nickelate in view of potential Ni-O bond disproportionation triggered by an onset of a SDW/CDW state.

## IV. SUMMARY AND DISCUSSION

In summary, we studied the electronic properties of the bilayer polymorph of $La_3Ni_2O_7$, a high-temperature superconductor under pressure, using the hybrid functional approach. We found a strong tendency for this system to develop localized magnetic moments on Ni ions and explored a variety of its magnetically ordered configu-rations, comparing their energetics and the associated microscopic charge distribution patterns.

TABLE II. Relative total energies and Ni-O distances in selected magnetic states of bilayer $La_3Ni_2O_7$ as obtained within HSE06 after full atomic position relaxation with the lattice parameters fixed to those measured experimentally at $T = 300$ K and $P = 1.6$ GPa.

| | Energy (meV/Ni) | $Ni\text{-}O^a$ (Å) | $Ni\text{-}O^p$ (Å) | $Ni\text{-}O^o$ (Å) |
|---|---|---|---|---|
| AFM-C | -17 | 1.97 | 1.93 | 2.20 |
| AFM-G | 0 | 2.00 | 1.93 | 2.24 |
| SDW AFM-G | -298 | 2.00/2.00 | 2.00/1.86 | 2.20/2.28 |
| SDW1 AFM-A | -319 | 2.00/2.00 | 1.99/1.87 | 2.19/2.29 |

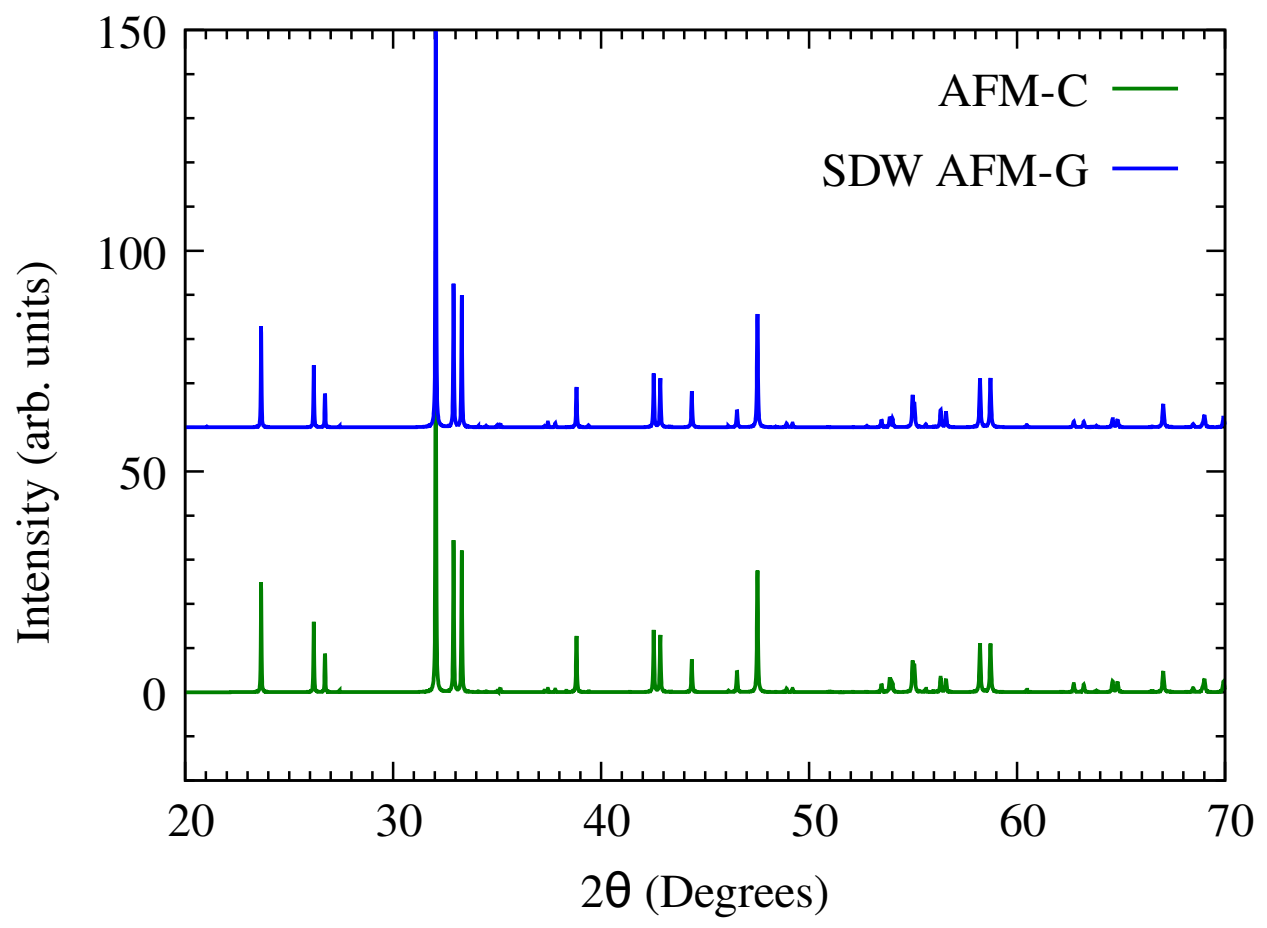

FIG. 11. X-ray diffractograms simulated for the high-symmetry AFM-C state and the symmetry-broken bond-disproportionated SDW AFM-G state, demonstrating barely noticeable differences.

We summarize some of the most important findings. First of all, the energy difference between the non-magnetic and magnetically ordered phases was found to be close to 0.75 eV per Ni, which corresponds to the Hund's rule coupling energy that appears with magnetic order but vanishes without magnetic order. Also of general importance is that all of the magnetic structures consistent with 4 Ni atoms per unit cell converged to energies in the range of 300 meV per Ni at most. What was really surprising to us is the exceptionally strong dependence of the low energy scale (*i. e.*, $\pm 1.5$ eV around $E_{\mathrm{F}}$) electronic structure on the type of magnetic order. Some magnetic structures were metallic, other semiconducting with gaps varying from about 0.5 to 1.2 eV. The $z$-projected magnetic moments varied between 1 and 1.5 $\mu_{\mathrm{B}}$, corresponding to a spin on Ni half way between 1/2 and 1, which is half way between what is expected for $Ni^{3+}$ ($S = 1/2$) and $Ni^{2+}$ ($S = 1$) if we consider a hole in only the $e_g$ symmetry crystal field split states.

Looking at the spin, orbital, and atomic projected densities of states, we conclude that the extra 0.5 holes per Ni beyond the base $Ni^{2+}$ $d^8$ configuration mainly occupy the O $2p$ states of either in-plane $x^2 - y^2$ symmetry about a particular Ni or, in some extreme cases, the inter-planar $O^{a}$ $2p_z$ orbitals. The latter was only found in those magnetic structures where the inter-planar spin coupling was ferromagnetic. These structures, however, also had considerably higher total energies than the magnetic structure with an antiferromagnetic inter-planar coupling. This is consistent with the exact diagonalization cluster calculations[60]. We conclude that bilayer $La_3Ni_2O_7$ quite strongly prefers not to have hole occupation of the inter-planar $O^{a}$ $2p_z$ orbitals in its lowest energy magnetic and charge density states.

We see strong tendencies towards the formation of molecular orbitals involving O atoms with high hole density. If the fifth hole is in the inter-planar O position, we have a strong ferromagnetic coupling due to double exchange and the formation of a 5-spin polaron. If, however, this O is fully occupied, which it is in all the low-energy cases, then the inter-planar exchange is antiferromagnetic, but, because of the low Ni-Ni coordination number of one, these $3z^2 - r^2$ orbitals will tend to form, via the intervening O, singlet states rather than to participate in an antiferromagnetic like spin structure. The singlet formation is facilitated by covalency effects that reduce the Hund's rule exchange tending to align the $3z^2 - r^2$ and $x^2 - y^2$ Ni spins. Unfortunately, a full description of the complex interplay of all these effects is beyond reach of either band theory or cluster calculations.

If we ordered the spins and charges in the system in a checkerboard fashion, we would have half of the Ni's involved in ZR singlets, with a remaining spin of 1/2 in a $3z^2 - r^2$ orbital, and the other half would be close to $Ni^{2+}$ with a spin of 1. This explains the 3 different kinds of spin and charge density wave states found in our hybrid functional calculations. The lowest energy 2 such CDW/SDW phases are of the AFM-A type, *i. e.*, ferromagnetic planes ordered antiferromagnetically. Qualitatively, this is also what is found in the exact diagonalization study[60] and corresponds to the ZRS – $Ni^{2+}$ checkerboard patterns in the two planes. These can be in phase, corresponding formally to the 6 hole – 4 hole cluster ordering, or out of phase, corresponding to the 5 hole – 5 hole ordering. This leads to the same kind of possible CDW/SDW structures that are found in the combined neutron powder diffraction (NPD) and muon-spin rotation/relaxation ($\mu$SR) study[64]. The total energy difference between these phases is very small of the order of 5 meV in hybrid functional calculations. However, we should note that, in spite of the energetic preference of the AFM-A CDW/SDW phases, the AFM-G SDW phase is the closest to the experimental magnetic moments in the NPD/$\mu$SR study[64]. The theoretical spin disproportionation is quite small for the AFM-A SDW phase compared to experiment.

We also studied the relaxed CDW/SDW structures and found that there was a considerable breathing mode like bond disproportionation in the phases with the ZRS like states. The Ni-O distance in the short bond was reduced by about 0.1 A, which is a very substantial bond disproportionation. Nonetheless, it is important that this bond disproportionation was all but invisible in the theoretical x-ray diffraction simulations. This kind of strong disproportionation is rather similar is size to that of $BaBiO_3$, which suggests a very strong electron-phonon coupling.

## V. APPENDIX A

A more general way of looking at the dimer problem has to do with the difference between treating things in terms of symmetry restricted single particle states as, say, in the trajectory from a molecular orbital approach or a Heitler-London (HL) approach to the $H_2$ molecule problem. In the HL approach for $U \gg 2t$, let us use perturbation theory but starting with two electrons, one on each atom with spin 1/2 and experiencing an effective antiferromagnetic coupling $J = 2t^2/U$ and developing eigenstates corresponding to one singlet and 3 degenerate triplets. Starting from the other limit $2t \gg U$, we treat $U$ as a perturbation and start with the molecular orbital limit, taking $U = 0$ and obtaining a bonding and anti-bonding combination of the two H 1$s$ orbitals split by $2t$ and a ground state with $S = 0$, *i. e.*, 2 antiparallel spin electrons in the bonding orbital which is a state that does not have a triplet associated with it. $S$ is identically zero, *i. e.*, the same as in the HL limit although the wave functions are very different. The excited anti-bonding state, if housing one of the two electrons at a cost of $2t$, could form a singlet or a triplet state with the electron in the bonding orbital. The singlet-triplet energy splitting would in perturbation theory be given by involving an effective $U^*$, which would effectively be between an electron in a bonding and the other in an anti-bonding orbital and would be given by the effective amount of double occupation of a site, which would be 1/4 and $U^* = U/4$. For the triplet state there can be no double occupation of a site but for the singlet state there can be, so it will be lowered by $(U/4)^2/2t$. This results in a ground state $S = 0$, and 4 excited states: 3 for the triplet and one for the singlet. Now, both the triplet and the singlet in the excited state have a state with $S_z = 0$ corresponding to the combinations $|\uparrow\downarrow\rangle + |\downarrow\uparrow\rangle$ for the triplet $m_s = 0$ and $|\uparrow\downarrow\rangle - |\downarrow\uparrow\rangle$ for the singlet. This points to a very small splitting in the excited state between the singlet and the triplet, much smaller than the splitting between the bonding and antibonding state of $2t$. Therefore, at zero or small magnetic field we retain an $S = 0$ ground state which, with its doubly occupied bonding state, has no triplet component because of the Pauli principle. This kind of an $S = 0$ state can be described in DFT obviously, but the kind of an $S = 0$ state that has a triplet component separated in energy by $J$ cannot be described in DFT because it is a linear combination of at least 2 Slater determinants distinguishing the triplet $S_z = 0$ state from the singlet $S_z = 0$ state.

So, it all depends on where we are in the HL vs. MO starting point scheme. If the splitting between the $d^7$ state and the $d^8\underline{L}$ state is small and the $t_{pd}$ is large compared to this splitting then we can better start with the band structure MO terminology. The splitting is given by $U_{dd} + J_{\rm H} - \Delta$. For the Nickelates like 2222 $La_3Ni_2O_7$, this energy difference is quite small and probably considerably less than 1 eV, but they, however, are likely not really in either limit. So, it is really interesting to look at both limits. However, as far as the dimer cluster is concerned, it basically solves the problem exactly, but only of the small cluster, while the band theory approach is forced to be in the MO limit since it involves only a single Slater determinant of symmetry-restricted wave functions.

## VI. APPENDIX B

We will compare the HSE06 with the LDA+U calculations for the AFM-C phase [panels (d) and (e) in Fig. 7]. In LDA+U, the following parameters were used: $U = 6$ and $J_{\rm H} = 0.7$ eV for Ni 3$d$ electrons and $U = 6$ eV for La 4$f$ electrons[54, 55, 70]. We see that although the PDOS between the two methods are closely related, there are important qualitative differences that stem from the inability of LDA+U to describe correlated molecular orbital states such as the $|\psi^{\mathbf{b}}\rangle$, $|\psi^{\mathbf{ab}}\rangle$, and $|\psi^{\mathbf{nb}}\rangle$ states of Ni dimers in $La_3Ni_2O_7$. This problem particularly affects MO states that are partially filled, and in the present case such are the $|\psi^{\mathbf{nb}}\rangle$ states. While HSE06 provides a clean separation $\Delta_{\mathrm{MO}}$ between the empty majority spin $|\psi^{\mathbf{ab}}\rangle$ states and the full majority spin $|\psi^{\mathbf{nb}}\rangle$ states, opening a charge gap of 1 eV [panel (d)], LDA+U mixes the anti-bonding and non-bonding molecular symmetries and fails to open the gap [panel (e)]. This is a universal issue of LDA+U that arises in correlated MO systems and was discussed in Ref. 58 in connection with the MO spin system $Na_2IrO_3$.

TABLE III. HSE06 relaxed atomic positions for the uniform AFM-C magnetic configuration.

| AFM-C at 1.6 GPa, space group: $Cmcm$ | | | |
|---|---|---|---|
| $a = 20.403$, $b = 5.3768$, $c = 5.4392$ Å, $\alpha = \beta = \gamma = 90°$ | | | |
| atom | $x$ | $y$ | $z$ |
| Ni | 0.0956 | 0.2470 | 0.25 |
| La1 | 0.1785 | 0.2573 | 0.75 |
| La2 | 0.5 | 0.2374 | 0.25 |
| O1 | 0.0848 | 0.5 | 0.5 |
| O2 | 0.0 | 0.1961 | 0.25 |
| O3 | 0.1073 | 0.0 | 0.0 |
| O4 | 0.2025 | 0.2908 | 0.25 |

TABLE IV. HSE06 relaxed atomic positions for the uniform AFM-G magnetic configuration.

| AFM-G at 1.6 GPa, space group: $Cmcm$ | | | |
|---|---|---|---|
| $a = 20.403$, $b = 5.3768$, $c = 5.4392$ Å, $\alpha = \beta = \gamma = 90°$ | | | |
| atom | $x$ | $y$ | $z$ |
| Ni | 0.0965 | 0.2460 | 0.25 |
| La1 | 0.1806 | 0.2604 | 0.75 |
| La2 | 0.5 | 0.2473 | 0.25 |
| O1 | 0.0868 | 0.5 | 0.5 |
| O2 | 0.0 | 0.1850 | 0.25 |
| O3 | 0.1099 | 0.0 | 0.0 |
| O4 | 0.2056 | 0.2945 | 0.25 |

TABLE V. HSE06 relaxed atomic positions for the bond-disproportionated SDW AFM-G magnetic configuration.

| SDW AFM-G at 1.6 GPa, space group: $Amm2$ | | | |
|---|---|---|---|
| $a = 5.4392$, $b = 20.403$, $c = 5.3768$ Å, $\alpha = \beta = \gamma = 90°$ | | | |
| atom | $x$ | $y$ | $z$ |
| Ni1 | 0.5 | 0.9032 | 0.3961 |
| Ni2 | 0.0 | 0.5963 | 0.4037 |
| La1 | 0.0 | 0.8196 | 0.4070 |
| La2 | 0.5 | 0.6806 | 0.3876 |
| La3 | 0.5 | 0.5 | 0.4031 |
| La4 | 0.0 | 0.0 | 0.4114 |
| O1 | 0.2422 | 0.4140 | 0.1613 |
| O2 | 0.5 | 0.0 | 0.3412 |
| O3 | 0.0 | 0.5 | 0.4728 |
| O4 | 0.7571 | 0.8897 | 0.1389 |
| O5 | 0.5 | 0.7963 | 0.4403 |
| O6 | 0.0 | 0.7070 | 0.3478 |

TABLE VI. HSE06 relaxed atomic positions for the bond-disproportionated SDW1 AFM-A magnetic configurations.

| SDW1 AFM-A at 1.6 GPa, space group: $Amm2$ | | | |
|---|---|---|---|
| $a = 5.4392$, $b = 20.403$, $c = 5.3768$ Å, $\alpha = \beta = \gamma = 90°$ | | | |
| atom | $x$ | $y$ | $z$ |
| Ni1 | 0.5 | 0.9033 | 0.3962 |
| Ni2 | 0.0 | 0.5961 | 0.4039 |
| La1 | 0.0 | 0.8196 | 0.4071 |
| La2 | 0.5 | 0.6807 | 0.3873 |
| La3 | 0.5 | 0.5 | 0.4030 |
| La4 | 0.0 | 0.0 | 0.4105 |
| O1 | 0.2433 | 0.4138 | 0.1598 |
| O2 | 0.5 | 0.0 | 0.3407 |
| O3 | 0.0 | 0.5 | 0.4726 |
| O4 | 0.7559 | 0.8898 | 0.1397 |
| O5 | 0.5 | 0.7963 | 0.4394 |
| O6 | 0.0 | 0.7072 | 0.3474 |

## VII. APPENDIX C

Tables III to VI present atomic positions of the four bilayer $La_3Ni_2O_7$ phases, AFM-C, AFM-G, SDW AFM-G, and SDW1 AFM-A, that were fully relaxed in HSE06 while keeping the lattice constants fixed to those measured experimentally at room temperature and low pressure of 1.6 GPa.

## VIII. ACKNOWLEDGMENTS

The authors would like to thank the Stewart Blusson Quantum Matter Institute and the Natural Sciences and Engineering Research Council of Canada for providing financial support and computational resources.